\documentclass[twocolumn,showpacs,preprintnumbers,amsmath,amssymb]{revtex4}

\usepackage{graphicx}
\usepackage{dcolumn}
\usepackage{bm}
\providecommand{\Journal}[4]{#1{\bf #2}, #3 (#4)}
\providecommand{\PL}{Phys. Lett. } %
\providecommand{\PLB}{Phys. Lett. B } %
\providecommand{\PR}{Phys. Rev. } %
\providecommand{\PRL}{Phys. Rev. Lett. } %
\providecommand{\PRC}{Phys. Rev. C } %
\providecommand{\PRD}{Phys. Rev. D } %
\providecommand{\RMP}{Rev. Mod. Phys. } %
\providecommand{\RMP}{Rev. Mod. Phys. } %
\providecommand{\PRN}{Prog. Part. Nucl. Phys. } %
\providecommand{\JPG}{J. Phys. G } %

\begin{document}

\preprint{Published in Chinese Physics Letters 25 (2008) 3920-3923.}

\title{Simple Classification of Light Baryons}

\author{Yan Chen}
\author{Bo-Qiang Ma}%
 \email{mabq@phy.pku.edu.cn}
\affiliation{ School of Physics and State Key Laboratory of Nuclear
Physics and Technology, Peking University, Beijing 100871, China}
\date{\today}

\begin{abstract}
We introduce a classification number $n$ which describes the baryon
mass information in a fuzzy manner. According to $n$ and $J^p$ of
baryons, we put all known light baryons in a simple table in which
some baryons with same ($n$, $J^p$) are classified as members of
known octets or decuplets. Meanwhile, we predict two new possible
octets.
\end{abstract}


\maketitle


The idea of the quark structure of hadrons appeared first in the
papers of Gell-Mann~\cite{Gell} and Zweig~\cite{Zweig}. It was shown
that the SU(3) octet symmetry can be realized on the basis of a
fundamental triplet of some hypothesized particles, called quarks by
Gell-Mann. At the beginning, quarks were take as a mathematical
expression of the SU(3) properties of hadrons, but soon it was
recognized that hadrons are bound systems of quarks: Baryon=$qqq$
and Meson=$q\bar{q}$. The light baryons are made up of  $u$, $d$,
$s$ quarks, which implies an approximate $SU(3)_{\mathrm{flavor}}$
symmetry. Up to now, a number of approaches have been developed for
describing the baryon mass spectrum, such as the SU(6)
model~\cite{Gursey, Sakita}, the bag model~\cite{Chodosa}, the
Skyrme model~\cite{Hayasgi140,M.P.Mattis}, the non-relativistic
quark model~\cite{Isgur1,Isgur5} and so on. These models incorporate
partly the dynamics of quantum chromodynamics~(QCD) and achieve
quantitative description. However, many predicted baryons of these
models have been experimentally observed while some baryons have not
been found. In addition, the predicted baryon masses of different
models are not the same, i.e., the predictions are model-dependent.
Although QCD has been widely accepted as a basic theory of strong
interaction, it is still a big challenge to compute the baryon mass
spectrum from first principle due to the complicated
non-perturbative and non-linear properties. Therefore a
comprehensive classification of all known baryons would be very
helpful for a fundamental understanding of strong interaction and
baryon structure.

In this Letter, we try to propose a simple method which directly
classifies the observed baryons to their possible multiplets
according to a fuzzy concept of mass range, rather than the precise
mass values. First, we introduce a classification number $n$ which
describes baryon mass range information. Secondly, according to $n$
and $J^p$, we put all light baryons into a table. We find that many
baryons with same $n$ and  $J^p$ can make up of a multiplet which is
an octet or a decuplet listed in the booklet of Particle Data
Group~(PDG)~\cite{PDG} based on the quark-model or in Ref.~\cite{GP}
based on SU(3) symmetry. We also predict two new possible octets.
The classification of known multiplets and the new predictions may
suggest the feasibility of the classification method although it is
phenomenological and rough.

We now introduce the classification number $n$, which is an integer
determined by two parameters. One is the center mass denoted as
$M_c^{B_{n}}$ for a kind of baryons with same isospin and
hypercharge, the other is the mass band width denoted as
$M_w^{B_{n}}$ corresponding to the center mass $M_c^{B_{n}}$. We
define $M_c^{B_{n+1}}-M_c^{B_{n}}=M_w^{B_{n}}$. If a baryon mass
$M_B$ satisfies
\begin{eqnarray}\label{eq1}
M_c^{B_{n}} - M_w^{B_{n-1}}/2 \leq M_B < M_c^{B_{n}} +
M_w^{B_{n}}/2,
\end{eqnarray}
the baryon belongs to a group of baryons with classification number
$n$.

To obtain reasonable $M_c^{B_{n}}$ and $M_w^{B_{n}}$, we briefly
review the relevant properties of the nonrelativistic potential
model~\cite{Rujula,Capstick}. The ground-state baryon mass is
\begin{eqnarray}\label{eq2}
M=\sum_{1}^{3}m_i +
\frac{2\alpha_s}{3}\frac{8\pi}{3}\left<\delta^3(\bm{r}_{ij})\right>
\sum_{i<j=1}^{3} \frac{\bm{S}_i \cdot \bm{S}_j}{m_i m_j},
\end{eqnarray}
where $m_i$ and $\bm{S}_i$ are the constituent quark effective mass
and spin of the $i$th quark, $\bm{r}_{ij}=\bm{r}_i-\bm{r}_j$, and
$\alpha_s$ is the quark-gluon coupling constant.
We suppose that the first center mass can be written as
$M_c^{B_{1}}=\sum_{1}^{3} m_i+m_0^B$, where $m_i$ is the $i$th
constituent quark effective mass and $m_0^B$ is a constant in
analogy with the second term of Eq.(\ref{eq2}). Using $m_{u,d}\sim
280$ MeV and $m_s \sim 440$ MeV, and also taking $m_0^B \sim$ 160,
150, 110, 50 MeV for hadrons $N/\Delta$, $\Lambda/\Sigma$, $\Xi$,
$\Omega$, we assume $M_c^{B_{1}}$s of $N/\Delta$, $\Lambda/\Sigma$,
$\Xi$, $\Omega$ being 1000, 1150, 1270 and 1370 MeV respectively.
The decrease of $m_0^B$ with the increase of the strange quark
number of a baryon is inspired by the $m_i m_j$ terms in the
denominators of Eq.(\ref{eq2}). However, we should take the above
values as from assumptions rather than from derivations. We need
also to mention that $\Delta 1$ and $\Omega 1$ are assumed to be
imaginary particles with spin $S=1/2$ for the sake of being
comparable with $N 1$, $\Lambda 1$, $\Sigma 1$, and $\Xi 1$.

The idea of the mass band width $M_w^{B_{n}}$ comes from the zeroth
order energies $E = (N + 3/2 )\omega =
(2n+l+3/2)\omega$~\cite{Capstick,Isgur5}, where $n$ is the number of
radial nodes, $l$ is the orbital angular momentum, and the
non-strange harmonic oscillator level spacing $\omega\simeq520$ MeV.
The mass band width $M_w^{B_{n}}$ is somewhat similar to $\omega$.
Therefore, we suppose that $M_w^{B_{n}}$ and $\omega$ are of the
same order. More explicitly, we suppose that $M_w^{B_{1}}$ is 400
MeV and $M_w^{B_{n^\prime }}$($n^\prime>1$) is 300 MeV.

\begin{table*}
\caption{The $n$ number and masses of  N, $\Delta$
baryons}\label{tab:1}
\begin{tabular}{|c|cccp{6mm}p{6mm}p{6mm}||c|p{11mm}p{11mm}p{12mm}p{14mm}p{8mm}p{7mm}p{7mm}|}
\hline $n_N$ & 1&   2 &  3  & 4  & 5&  ~ 6 &  $n_\Delta$ & ~ 1&   ~2&  ~3  & ~4  & ~5 &~6&~7 \\
\hline
$M_c^{N}$(MeV) &1000& 1400 &1700 &2000& 2300
&2600&$M_c^{\Delta}$(MeV)  &1000 &1400&1700    &2000 &2300
&2600&2900\\\hline
$J^+$&      &  N  &   masses  &      &     &&$J^+$    &        &&$\Delta$&masses&&&\\
\hline
1/2&   939 &1440 & 1710 & 2100 &     &     &1/2&       &        & 1750   & 1910&     &&          \\
3/2 &       &     &1720  &1900 &     &     &3/2&       & 1232   & 1600   & 1920&     &&          \\
5/2 &       &     &1680  &2000 &     &     &5/2&       &        &        & 1905/2000 &        && \\
7/2 &       &     &      &1990 &     &     &7/2&       &        &        & 1950      &2390    &&\\
9/2 &       &     &      &     & 2220&     &9/2&       &        &        &           &2300    &&\\
11/2&       &     &      &     &     &     &11/2       &        &        &           &        &2420&  &\\
13/2&       &     &      &     &     &2700 &13/2&      &        &        &           &        &&\\
15/2&       &     &      &     &     &     &15/2&      &        &        &           &        &&2950\\
\hline $J^-$&   & && & &&  $J^-$ & &&&&&&\\\hline
1/2 &     &1535  &1650  &2090&       &     &1/2&       &  & 1620 & 1900  &2150  && \\
3/2 &     &1520  &1700  &2080&       &     &3/2&       &  & 1700 & 1940  &      &&\\
5/2 &     &      &1675      &    &2200   &     &5/2&       &  &      & 1930  &2350  && \\
7/2 &     &      &      &    &2190   &     &7/2&       &  &      &       &2200  &&\\
9/2 &     &      &      &    &2250   &     &9/2&       &  &      &       &2400  && \\
11/2&     &      &      &    &       &2600 &11/2&      &  &      &       &&&       \\
13/2&     &      &      &    &       &     &13/2&      &  &      &       &&& 2750 \\
\hline
\end{tabular}
\end{table*}

\begin{table*}
\caption{The $n$ number and masses of  $\Lambda$, $\Sigma$ baryons
}\label{tab:2}
\begin{tabular}{|c|ccccc||c|ccccccc|}
\hline $n_\Lambda$ &  1&   2 &  3  & 4  & 5  & $n_\Sigma$ &  1&   2 &  3  & 4   &5 &6&7\\
\hline
$M_c^{\Lambda}$(MeV) &1150 &1550&1850    &2150    &2450& $M_c^{\Sigma}$(MeV)&1150 &1550&1850    &2150   & 2450&2750&3050\\
\hline $J^+$&&     $\Lambda$    &     masses   &&&$J^+$&       &$\Sigma$&masses&& &&\\
\hline
1/2 &  1116 &1600    & 1810 &                &     &1/2&  1193 &1660     &1880/1770                   &          &&&\\
3/2 &       &        & 1890 &\underline{2000}&     &3/2&       &1385/\underline{1560}/\underline{1690}&1840&2080 &&&  \\
5/2 &       &        & 1820 & 2110&                &5/2&       &         &1915                             &2070 &&& \\
7/2 &       &        &      & 2020&                &7/2&       &         &                                 &2030 &&&\\
9/2 &       &        &      &     & 2350           &9/2&       &         &          &                      &\underline{ 2455 }&&\\
11/2&       &        &      &     &\underline{2585}&11/2&      &         &          &                      &&&\\
15/2&       &        &      &     &                &15/2&      &         &          &                      &&&\underline{3170}\\
\hline
$J^-$&   & &&&&$J^-$&   & &&&&&\\
\hline
1/2 &&1405/1670  &1800  &     &     &1/2 &     &  1620                     &1750&2000            &&&\\
3/2 &&1520/1690  &      &     &2325 &3/2 &     & 1670/1580/\underline{1480}&1940&\underline{2250}&&&\\
5/2 & &          &1830  &     &     &5/2 &     &                           &1775&                &&&\\
7/2 & &          &      &2100 &     &7/2 &     &                           &    &2100            &&&\\
11/2& &          &      &     &     &11/2&     &                           &    &                &&\underline{2620}&\\
13/2& &          &      &     &     &13/2&     &                           &    &                &&&\underline{3000}\\
\hline
\end{tabular}
\end{table*}

\begin{table}
\caption{The $n$ number and masses of  $\Xi$, $\Omega$
baryons}\label{tab:3}
\begin{tabular}{|c|ccccc|}
\hline $n_\Xi$&  1&   2 &  3  & 4  & 5   \\
\hline
$M_c^\Xi$(MeV) &1270 &1670&1970    &2270    &2570 \\
\hline $J^+$&&   $\Xi$     &  masses      & &\\\hline
1/2 &  1318 & \underline{1690 }  &          && \\
3/2 &       & 1530   &         &\underline{2250} & \\
5/2 &       &        &\underline{2030}     & &     \\
7/2 &       &        &         &\underline{2120} &     \\
\hline
$J^-$&   & &&&\\
\hline
1/2 &     & \underline{ 1620}&  &     &    \\
3/2 &     & 1820 &  & \underline{2370}  &     \\
5/2 &     &  &  \underline{1950 }     &    &     \\
7/2 &     &  &          &      & \underline{2500 }  \\
\hline
$M_c^\Omega$(MeV) &1370 &1770&2070    &2370    &2670 \\
\hline $J^+$&&    $\Omega$     &  masses      & &\\\hline
3/2 &       & 1672  &    & \underline{2470}&        \\
5/2 &       &        &        & \underline{2380}&      \\
7/2 &       &        &        & \underline{2250}&      \\
\hline
\end{tabular}
\end{table}

Using the relation of baryon masses and the classification number
$n$, i.e.,  Eq.~(\ref{eq1}), we put baryons N, $\Lambda$, $\Sigma$,
$\Xi$, $\Delta$ and $\Omega$ in Tables~\ref {tab:1}, \ref{tab:2} and
\ref{tab:3} respectively. The underlined baryons mean that their
spins and parities are not confirmed yet. We try to put them in
these tables according to their masses. It is noted that an
underlined baryon might be putted in another place in these tables
as long as its mass satisfies Eq.~(\ref{eq1}).

\begin{table*}
\caption{Mass table of light baryons classified by ($n$,$J^P$) (in
units of MeV) } \label{tab:all}
\begin{ruledtabular}
\begin{tabular}{||c|p{15mm}|rccccp{10mm}|p{7mm}|p{9mm}|l||}
Group&\hspace{-10pt}(n,~$J^p$)& N &
$\Lambda$&$\Sigma$&$\Xi$&$\Delta$&$\Omega$ &\hspace{-14pt}Singlet&\hspace{-10pt}Octet&\hspace{-14pt}Decuplet\\
&&&
&&&&&\hspace{-4pt}$\Lambda$&\hspace{-10pt}N$\Lambda$$\Sigma$$\Xi$&\hspace{-10pt}$\Delta$$\Sigma$$\Xi$$\Omega$\\
\hline
1&$\hspace{-10pt}(1,1/2^+)$ & 939  &1116& 1193& 1318 &  & && $\star$&\\
2&$\hspace{-10pt}(2,1/2^+)$ & 1440 &1600& 1660& \underline{1690}& && & $\ast$&\\
3&$\hspace{-10pt}(3,1/2^+)$ & 1710 &1810&1880& & && & $\star$&\\
4&$\hspace{-10pt}(3,1/2^+)$ &      &    &1770& &1750 && & ?&?\\
5&$\hspace{-10pt}(4,1/2^+)$ &2100  &    &    & &1910 & &&? &?\\
6&$\hspace{-10pt}(2,3/2^+)$ &  &&1385& 1530& 1232&1672 & &&$\star$\\
7&$\hspace{-10pt}(2,3/2^+)$ & &&\underline{1560}& && & &? &?\\
8&$\hspace{-10pt}(2,3/2^+)$ & &&\underline{1690}& && & &? &?\\
9&$\hspace{-10pt}(3,3/2^+)$ & 1720 &1890&1840& & 1600& && $\ast$&$\star$\\
10&$\hspace{-10pt}(4,3/2^+)$ & 1900 &\underline{2000}&2080& & 1920& \underline{2470}&& $\odot$&$\ast$\\
11&$\hspace{-10pt}(3,5/2^+)$ & 1680 &1820&1915& 2030& && &$\star$ &\\
12&$\hspace{-10pt}(4,5/2^+)$ & 2000 &2110&2070& \underline{2250}&1905&\underline{2380}& &$\odot$ &$\ast$\\
13&$\hspace{-10pt}(4,5/2^+)$ &      &    &     &       &2000& &  & &?\\
14&$\hspace{-10pt}(4,7/2^+)$ & 1990 &2020&2030& \underline{2120}& 1950&\underline{2250}&& ?&$\ast$\\
15&$\hspace{-10pt}(5,7/2^+)$ &      &    &    &     & 2390& & &&?\\
16&$\hspace{-10pt}(5,9/2^+)$ & 2220 &2350&\underline{2455}& & 2300& && $\star$&?\\
17&$\hspace{-10pt}(5,11/2^+)$ &     & \underline{2585}  &   & &2420& &?& ?&$\star$\\
18&$\hspace{-10pt}(6,13/2^+)$ &2700   &    &    & && &&? &\\
19&$\hspace{-10pt}(7,15/2^+)$ &   &    &   \underline{3170} & &2950& && ?&?\\
\hline
20&$\hspace{-10pt}(2,1/2^-)$ & 1535 &1670& 1620&  &1620& && $\star$&$\star$\\
21&$\hspace{-10pt}(2,1/2^-)$ &  &1405& &  && &  $\star$&&\\
22&$\hspace{-10pt}(3,1/2^-)$ & 1650 &1800&1750& & & && $\star$&\\
23&$\hspace{-10pt}(4,1/2^-)$ & 2090 &&2000& & 1900& && ?&?\\
24&$\hspace{-10pt}(5,1/2^-)$ & &&& & 2150& && &?\\
25&$\hspace{-10pt}(2,3/2^-)$ & 1520 &1690&1670& 1820& & && $\star$&\\
26&$\hspace{-10pt}(2,3/2^-)$ & &1520 &1580& & && $\star$ & ?&?\\
27&$\hspace{-10pt}(2,3/2^-)$ & & &\underline{1480}&\underline{1620} & && & ?&?\\
28&$\hspace{-10pt}(3,3/2^-)$ & 1700 &&1940& & 1700& &&$\ast$&$\star$\\
29&$\hspace{-10pt}(4,3/2^-)$ & 2080 &&\underline{2250}& \underline{2370}& 1940&& & ?&?\\
30&$\hspace{-10pt}(5,3/2^-)$ &  &2325&& & & &?&? &\\
31&$\hspace{-10pt}(3,5/2^-)$ & 1675 &1830&1775& \underline{1950}&&& &$\ast$ &\\
32&$\hspace{-10pt}(4,5/2^-)$ &  &&& &1930& && &?\\
33&$\hspace{-10pt}(5,5/2^-)$ & 2200 &&& & 2350& && ?&?\\
34&$\hspace{-10pt}(4,7/2^-)$ &  &2100 & 2100 & & && & ?&?\\
35&$\hspace{-10pt}(5,7/2^-)$ & 2190 &&&\underline{2500} &2200 && & ?&?\\
36&$\hspace{-10pt}(5,9/2^-)$ & 2250 &&& & 2400& && $\star$&?\\
37&$\hspace{-10pt}(6,11/2^-)$ & 2600 && \underline{2620}& && && ?&?\\
38&$\hspace{-10pt}(7,13/2^-)$ &  &&\underline{3000}& &2750& && ?&?\\
\end{tabular}
\end{ruledtabular}
{\footnotesize The underlined baryons mean that their spins and
parities are unknown and we put them in the table according to their
classification numbers $n$ artificially. $\star$ means that the
baryons in the group belong to an octet or a decuplet listed in PDG,
$\ast$ means the baryons belong to a multiplet listed in
Ref.~\cite{GP}, and $\odot$ means that the baryons belong to our
prediction of a new octet, and $?$ means that the baryons are still
unknown to a multiplet.}
\end{table*}

 Then according to $n$ and $J^p$, we put all light
baryons listed in PDG~\cite{PDG} in Table~\ref{tab:all}. In this
table, there are 38 groups with same $n$ and $J^p$, noting that
baryons $\Sigma$, $\Xi$ in one group may belong to an octet or a
decuplet. We find that many observed baryons with same $n$ and $J^p$
are directly classified to their possible multiplets, thirteen of
which are the same with the multiplets in PDG~\cite{PDG} and seven
of which are the same with the multiplets in Ref.~\cite{GP}. There
are also some baryons which can not be classified to a multiplet
clearly.

From Table \ref{tab:all}, we predict two new possible octets marked
with $\odot$. One is (N1900, $\Lambda$2000, $\Sigma$, $\Xi$) of
$J^p=3/2^+$, the other is (N2000, $\Lambda$2110,$\Sigma$, $\Xi$) of
$J^p=5/2^+$. The $\Sigma$ and $\Xi$ mass ranges of these two octets
are 2000-2300 MeV and 2100-2400 MeV respectively. We calculate the
baryon decay widths to check our prediction.

For the decay process of a baryon B$^*$  to a baryon B and a
pseudoscalar meson M
\begin{equation}B^* \to B + M ,\end{equation}
the calculation of decay widths can be performed in the framework of
Rarita-Schwinger formalism.

The parity-conserving Lagrangian and decay widths of the process
$B^*_{3/2^+} \rightarrow B_{1/2^+} + M$ are~\cite{JGR}
\begin{eqnarray}
\mathcal{L}&=& \frac{g_{B^*BM}}{m_{\pi}} \bar{\Psi}\Phi^{\mu_1}\partial_{\mu_1}\phi,\\
\Gamma &=&  \frac
{g^2_{B^{*}BM}P^3_{cm}[(m^*_B+m_B)^2-m^2]}{24\pi(m_1m_\pi)^2}.
\end{eqnarray}
Those of the process $B^*_{5/2^+} \rightarrow B_{1/2^+} + M$ are
\begin{eqnarray}
\mathcal{L}&=&i\frac{g_{B^*BM}}{m^2_{\pi}}
\bar{\Psi}\gamma^5\Phi^{\mu_1\mu_2}\partial_{\mu_1}\partial_{\mu_2}\phi,\\
\Gamma &=&  \frac
{g^2_{B^{*}BM}P^5_{cm}[(m^*_B-m_B)^2-m^2]}{30\pi(m_1m^2_\pi)^2},
\end{eqnarray}
where
\begin{eqnarray}
P_\mathrm{cm} =
\frac{\{[m^{*2}_B-(m_B+m)^2][m^{*2}_B-(m_B-m)^2]\}^{1/2}}{2m^*_B},
\end{eqnarray}
with $P_\mathrm{cm}$ being the c.m. momentum of final particles,
$g_{B^{*}BM}$ being the universal coupling constant~\cite{SGM},
$m_{B}^{*}$ and $m_B$ being the baryons masses, and $m$ being the
meson mass. The results are listed in Table~\ref{tab:tab5}, from
which we notice that  most predicted decay widths are consistent
with the experimental data.

\begin{table*}
\caption{\label{tab:tab5}The masses and widths of baryons (in units
of MeV)}
\begin{ruledtabular}
\begin{tabular}{cccccc}
 & PDG width & Decay mode & Branching ratio &$\Gamma_{i}(\mathrm{exp})$  & $\Gamma_{i}(\mathrm{th})$  \\
\hline
$N(1900)$ & 420-576 & $N\pi$ &  20\%-32\% & 84.0-184.3 &110.0\\
& & $N\eta$ & 9\%-19\% & 37.8-109.4 & 38.2 \\
& & $\Lambda K$ & 2.10\%-2.70\%& 8.8-15.5  & 4.8 \\
$\Lambda(2000)$ & 80-180 & $\sqrt{\Gamma_{N\bar{K}}\Gamma_{\Sigma\pi}}$ & $-24$\%-$-16$\% & $-43.2$-$-12.8$ &  $-58.0$\\
&&&&$\alpha=-0.20$&$A_8$=10.5\\
\hline
$N(2000)$ & 180-800 & $N\pi$ &  3\%-13\% & 5.4-104.0 &73.1\\
$\Lambda(2110)$ & 150-250& $N\bar{K}$ & 5\%-25\% & 7.5-62.5 & 10.3\\
& & $\Sigma\pi$ &10\%-40\% & 15.0-100.0 & 85.3\\
&&&&$\alpha=-0.10$&$A_8$=4.0\\
\end{tabular}
\end{ruledtabular}
$\alpha$, $A_8$ are the parameters of the universal coupling
constants for the $\textbf8 \to \textbf8+\textbf8$ decays.
\end{table*}

However, in Table~\ref{tab:all}, there is a puzzle, i.e., several
baryons of a same kind may have the same $n$ and  $J^p$.
There are 5 such cases, which are groups 3 and 4, groups 6, 7 and 8,
groups 12 and 13, groups 20 and 21, and groups 25, 26 and 27. There
are several possible solutions. Maybe when more baryons are
observed, baryons with the same $n$ and $J^p$ may belong to
different multiplets. It was suggested in Ref.~\cite{Azimov} that
$\Sigma$(1480) and $\Xi$(1620) might be members of a new light
octet, whose N member is predicted to have the mass around 1100 MeV
and the vanishingly small total width. Maybe some baryons have the
same $n$ and  $J^p$ can mix with each other, for example, the two
$\Lambda$'s of the groups 20 and 21 may mix, and so do the
$\Lambda$'s of the groups 25 and 26. Maybe some baryons belong to a
multiplet with exotic baryons. For example, $\Sigma$(1770) makes it
a potential candidate for the $\Sigma_{\overline{10}}$ member of the
antidecuplet~\cite{GP}. Four groups of baryons $\Sigma$(1560),
$\Sigma$(1690), $\Sigma$(1580), $\Delta$(2000) have no solution and
need more study.

Although the introduced center mass $M_c^{B_{n}}$ and mass band
width $M_w^{B_{n}}$ are arbitrary in some sense, our simple method
can directly classify the observed baryons to their possible
multiplets known in the literatures. The classification of known
multiplets and predictions of new possible multiplets may suggest
the feasibility of the introduction of the classification number $n$
which is based on the fuzzy concept of mass range instead of exact
mass values. The simple classification of all known light baryons in
Table~IV might be inspiring for both experimental and theoretical
studies: experimentalists may search for possible missing baryons by
looking at vacancies in the table and theorists may seek for better
classification schemes of baryons and reveal more fundamental
relations between different baryons. Of course, our simple method
should be only considered as a roughly phenomenological attempt to
classify all light baryons. A more comprehensive and elegant
classification should be searched for from more fundamental and
profound considerations.

We are grateful to Qihua Zhou and Bin Wu for useful discussions.
This work is partially supported by National Natural Science
Foundation of China (Nos.~10421503, 10575003, 10528510), by the Key
Grant Project of Chinese Ministry of Education (No.~305001), by the
Research Fund for the Doctoral Program of Higher Education (China).

\end{document}